\documentclass[12pt,a4paper]{iopart} 
\usepackage{bm}
\usepackage{comment}
\usepackage{graphicx}
\usepackage{amsfonts}
\usepackage{tikz}
\usepackage{siunitx}
\usepackage{longtable}
\usepackage{dcolumn}
\usepackage{cite}
\newcolumntype{d}[1]{D{.}{.}{#1}}
\newcommand\mc[1]{\multicolumn{1}{c}{#1}} 

\newcommand{\tj}[6]{\left({#1\atop#4}\,{#2\atop#5}\,{#3\atop#6}\right)} 

\newcommand{\red}[1]{{\textcolor{red}{#1}}}

\begin{document}

\title[B-Spline basis Hartree-Fock method]
{B-Spline basis Hartree-Fock method for arbitrary central potentials: atoms, clusters and electron gas}

\author{D T Waide, D G Green and G F Gribakin}

\address{School of Mathematics and Physics, Queen's University Belfast, University Road, Belfast BT7 1NN, Northern Ireland, UK}

\eads{\mailto{dwaide01@qub.ac.uk}, \mailto{d.green@qub.ac.uk}, \mailto{g.gribakin@qub.ac.uk}}

\begin{abstract}
An implementation of the Hartree-Fock (HF) method capable of robust convergence for well-behaved arbitrary central potentials is presented. The Hartree-Fock equations are converted to a generalized eigenvalue problem by employing a B-spline basis in a finite-size box. Convergence of the self-consistency iterations for the occupied electron orbitals is achieved by increasing the magnitude of the electron-electron Coulomb interaction gradually to its true value. For the Coulomb central potential, convergence patterns and energies are presented for a selection of atoms and negative ions, and are benchmarked against existing calculations. The present approach is also tested by calculating the ground states for an electron gas confined by a harmonic potential and also by that of uniformly charged sphere (the jellium model of alkali-metal clusters). For the harmonically confined electron-gas problem, comparisons are made with the Thomas-Fermi method and its accurate asymptotic analytical solution, with close agreement found for the electron energy and density for large electron numbers. We test the accuracy and effective completeness of the excited state manifolds by calculating the static dipole polarizabilities at the HF level and using the Random-Phase Approximation. Using the latter is crucial for the electron-gas and cluster models, where the effect of electron screening is very important. Comparisons are made for with experimental data for sodium clusters of up to $\sim $100 atoms.
\end{abstract}





\section{Introduction}
Much of computational and theoretical atomic physics and chemistry relies on accurate electronic structure calculations.
The general $N$-electron problem is, however, intractable. 
A favoured starting point for electronic structure calculations of atoms, molecules and nanostructures is the Hartree-Fock (HF) approximation \cite{hartree_wave_1928,slater_note_1930,fock_naherungsmethode_1930}.  
This self-consistent-field approximation treats each electron as if it were moving in the field of the nuclei and the mean field of all the electrons.
This can provide a good first approximation and even be sufficient in simple problems. Moreover, the corresponding electronic states can also be used a basis for higher-order calculations that account for post-HF correlations, e.g., using many-body theory (see e.g., \cite{boylepindzola,Safronova:MBT:elcAtom,Gribakin:2004,DGG_posnobles,DGG:2018:PRL}). 

Convergence of methods which aim to achieve self-consistency by iterations is not guaranteed. Recent investigation has shown that convergence of Hartree-Fock-like problems exhibits a fractal nature based on the choice of parameters used \cite{theel_fractal_2017}. This makes it impossible to predict whether or not the method will converge for a given system. Special measures need to be taken to ensure that iterations for a large class of systems converge and that they converge quickly.
Here, we use a numerical implementation of the self-consistent Hartree-Fock equations based on a B-spline approach and a new algorithm for achieving self-consistency ({\tt BSHF}) \cite{WaideCPC} that allows the use of arbitrary central potentials. Specifically, convergence is aided by gradually increasing the magnitude of the electron-electron Coulomb interaction to its true value.
We show that it provides good convergence properties for difficult systems such as negative ions, or electrons confined by a weak harmonic potential \cite{ahlrichs_hartree-fock_1975} (for which traditional convergence methods failed, and for which we compare with the results of the Thomas-Fermi method). 
We calculate static dipole polarizabilities at the HF level and using the Random-Phase Approximation (RPA) to assess the degree of completeness of the excited state manifold. This understanding is important for future applications, such as studying the interaction of a positron and electrons in a harmonically confined electron gas \cite{makkonen_enhancement_2014} using many-body theory methods \cite{gribakin_many-body_2004,green_positron_2014}.




\section{Hartree-Fock method and its present B-spline basis numerical implementation}
The relevant HF equations and the numerical implementation used here are described fully in Ref. \cite{WaideCPC}. 
In the Hartree-Fock approximation \cite{hartree_wave_1928,slater_note_1930,fock_naherungsmethode_1930}, the total wavefunction of an $N$-electron system of energy $E$ is approximated by a Slater determinant (or sum of Slater determinants, in general) that is an antisymmetrised product of $N$ single-electron spin orbitals $\phi_{\alpha_j}(x_j)$, viz.~$\Psi_E(x_1,\dots,x_N)= \sqrt{N!}\,\hat{\mathcal{A}}~\prod_{j=1}^{N} \phi_{\alpha_j}(x_j)$, 
where 
$\mathcal{A}$ is the antisymmetrisation operator,
$\alpha_j$ represents a complete set of quantum numbers describing the $j$-th orbital, and $x_j=({\bf r}_j,\sigma _j)$ represents the electron position and spin.
Minimising the expectation value of the Hamiltonian through the variation of the $\phi_\alpha$, or, in the diagrammatic approach, summing a certain class of diagrams \cite{Goldstone:1957,mbtexposed}, yields the system of $N$ integro-differential equations, the Hartree-Fock equations, for the electron orbitals $\phi_{\alpha_j}\equiv \phi _j$ and single-particle energies $\varepsilon_j$, in atomic units,
\begin{eqnarray} \label{eqn:HF}
\left(-\frac{1}{2}\nabla^2 + V(r) + \hat{V}^{\rm HF}\right)\phi_j(x) = \varepsilon_{j} \phi_{j}(x).
\end{eqnarray}
Here the first term in the bracket is the kinetic energy operator, and the second term is a local central potential $V(r)$. For an atom with atomic number $Z$, $V(r)=-Z/r$, but in the {\tt BSHF} program  \cite{WaideCPC} it can also be chosen to be an arbitrary central potential, e.g., a harmonic confining potential, for a system of electrons to approximate the electron gas in the background of a uniform positive-charge distribution \cite{Zubiaga2016}. 
The Hartree-Fock potential $\hat{V}^{\rm HF} = \sum_{i=1}^{N}\left(\hat{J}_{i} - \hat{K}_{i}\right)$ is a sum of the direct and (non-local) exchange terms 
$\hat{J}_i \phi_j (x) =  \int dx_i~\phi^*_i(x') \rho ^{-1} \phi_i(x') \phi_{j}(x)$ 
and
$\hat{K}_i \phi_j (x) =  \int dx_i~\phi^*_i(x') \rho ^{-1} \phi_j(x') \phi_{i}(x)$,
where $\rho = |{\bf r}' - {\bf r}|$.
Equation (\ref{eqn:HF}) demands a self-consistent solution due to the interdependency of the Coulomb mean field potential $\hat V^{\rm HF}$ and the electron orbitals $\phi _j$.
Beyond this, the resulting ground-state orbitals can be held \textit{frozen}, i.e. constant, while an additional electron or positron is introduced and the wavefunction of this extra particle in the presence of the \textit{frozen core} is subsequently found, enabling generation of excited-state bases for higher-order many-body calculations. 

For a spherical system confined by a central potential $V(r)$, single-particle wavefunctions can be written in terms of their radial, angular and spin components,
\begin{equation}
\phi_{nlm\sigma} (x)= r^{-1}P_{nl}(r)Y_{lm}(\theta,\phi)\chi_\sigma \,,
\end{equation}
where $P_{nl}$ is the radial wavefunction, $Y_{lm}$ is the spherical harmonic and $\chi_\sigma$ is the spin part.  

In the case of a closed-shell system, each spatial electronic orbital is occupied by two electrons with antiparallel spins. For a closed-shell atom or another spherical system, the solution of the HF equations reduces to a \textit{central-field problem}, i.e. the potential is spherically symmetric, leading to the radial Hartree-Fock equations,
\begin{equation}\label{eqn:hf-1}
\fl -\frac12\frac{\rmd ^2}{\rmd r^2}P_i(r) + \left[V(r)+\frac{l(l+1)}{2r^2}+V_\text{dir}(r)\right] P_i(r) 
+ \int_{0}^{\infty} U(r,r')P_i(r') \rmd r' = E_iP_i(r),
\end{equation}
where $i$ is a composite label for the orbitals of quantum numbers $nl$. Here $V_\text{dir}(r)$ is the local direct (Hartree) potential, representing the interaction between a single particle and the average field of the other particles, and $U(r,r')$ is the non-local exchange kernel (the ``Fock term''), which is the interaction between two particles due to the exchange of their coordinates (see \cite{WaideCPC} for detailed expressions). 

Equation \eref{eqn:hf-1} may also be used for open-shell electronic configurations under the further approximation of spherical averaging of the wavefunctions. In the case of the ground state for open-shell neutral atoms this difference is relatively small since the variation from the ideal case arises due to only one incomplete subshell. 

Wavefunctions of excited electron states can also be obtained from Eq.~(\ref{eqn:hf-1}). Physically, they describe an electron added to the ground-state of $N$ electrons. In the case of an additional electron in a system with $s$ ground-state orbitals, the wavefunction of the additional particle, $i>s$, is found from Eq.~(\ref{eqn:hf-1}) with
$V_\text{dir}(r)$ and $U(r,r')$ fixed by the self-consistent ground-state calculation of the first $s$ orbitals. For positrons, one simply drops the exchange term and makes appropriate changes of sign to account for its opposite charge.


\subsection{Hartree-Fock equation in the B-spline basis}
B-splines $B_{i,k}$ of order $k$ are piecewise polynomials of degree $k-1$ defined over a restricted domain (``box'') that is divided into $n-k+1$ segments by a knot sequence of $n$ points: $r_i\in[0, R]$, where $i=1,\dots,n$ and $n$ is the number of non-zero splines in the basis \cite{de-boor}.
We use an exponential knot sequence to ensure that the wavefunctions are represented accurately near the nucleus where they vary more rapidly, and to minimise computational expense in regions far from the nucleus where the wavefunction varies least rapidly. Such knot sequences also generate sets of excited states that ensure rapid convergence of perturbation-theory sums and are useful in many-body theory calculations \cite{gribakin_many-body_2004}. By varying the parameters of the sequence, one can make it almost equidistant, which is useful for studying a harmonically confined electron gas and models of clusters.

The radial wavefunction is expanded in terms of B-splines, $P_i(r) = \sum_j c^{(i)}_j B_j(r)$, where we have dropped the second subscript in the B-splines. Projecting the radial Hartree-Fock equation for a given angular momentum $l$ onto an arbitrary basis function $B_k(r)$ gives 
%
the generalized eigenvalue problem,  ${\mathcal{\bm H}}^{(l)} {\bm c}^{(l)} = E \mathcal{B}{\bm c}^{(l)}$, where
\begin{equation}\label{eqn:h-mat-elem} 
\mathcal{H}^{(l)}_{ij} = \int_0^R \left\{
\frac{1}{2}\frac{\rmd B_i}{\rmd r}\frac{\rmd B_j}{\rmd r}
+ B_i(r) \left[V(r) + \frac{l(l+1)}{2r^2}+\hat V_\mathrm{HF}\right] B_j(r)\right\} \rmd r,
\end{equation}
$\mathcal{B}_{ij}=\langle B_i|B_j\rangle$ is the B-spline overlap matrix, ${\bm c}^{(l)}$ is the vector of coefficients for angular momentum $l$, $V(r)$ is the local central potential, and $\hat V_\mathrm{HF}$ is  the Hartree-Fock potential.  Note that to implement the boundary conditions $P(0) = P(R) = 0$, the first and last splines are discarded in the expansions. The eigenvalue problem is computationally easier to solve than the original set of integro-differential equations. The integrals are calculated using Gauss-Legendre quadrature, splitting the integration interval into two sections in the elements where a cusp is present \cite{WaideCPC}.

\subsection{Convergence of iterations}
In our approach, the electron potential is calculated, starting with the wavefunctions in the potential $V(r)$ (i.e., $\hat V_{\mathrm{HF}}=0$), and the eigenvalue problem is solved repeatedly until the difference in successive approximations of the wavefunction and energy decreases below a certain tolerance $\varepsilon$, i.e., when $\eta_i<\varepsilon$ for all orbitals, where $\eta_i$ is defined as
\begin{equation}
\eta_i = \mathrm{max}\left( \left| P_i^{(m)}(r)-P_i^{(m+1)}(r)\right|,
\left| E_i^{(m)}-E_i^{(m+1)}\right| \right).
\end{equation}
The HF problem is initially solved only for values of orbital angular momentum represented by the occupied orbitals. 
Once the solutions of the occupied orbitals are self-consistent to tolerance $\varepsilon$, the excited electron or positron states are calculated for all required values of the angular momentum.

To ensure that the solutions converge towards the true ground state of the many-electron system and do not get `stuck' on local minima representing excited states or fail to converge, being caught in a cycle of distinct solutions, it is neccessary to employ a number of strategies.
In particular, the strength of the elementary charge $e$ is initially chosen to be smaller than unity so that the effect of the electron-electron interaction on the solution to the HF equations is suppressed. The HF problem is solved using this reduced interaction. Once self-consistent solutions have been determined, the strength of the elementary charge used in the electron-electron interaction, $e_m$, is increased towards unity according to
\begin{equation}\label{eq:anneal}
e_m = 1-\frac12 e_{\rm s}^m,\quad m=1,\dots,m_\mathrm{max},
\end{equation}
where typically $e_{\rm s}\sim0.15$--0.5, and $m_\mathrm{max}$ is the number of increments to take before setting the charge to unity, and iterating further until self-consistency is reached.

To improve the rate at which the solution converges, in particular, when iterations are trapped in a cycle, the new wavefunctions may be calculated as a linear combination of the current iteration and the previous one, viz.,
\begin{equation}\label{eq:wf_mix}
P_{i,est}^{(m)}(r) = (1-\alpha)P_i^{(m)}(r) + \alpha P_i^{(m-1)}(r),
\end{equation}
where the coefficient $\alpha$ is calculated as
\begin{equation}
\alpha = \frac{E_i^{(m)}-E_i^{(m-1)}}{E_i^{(m)}-2E_i^{(m-1)}+E_i^{(m-2)}},
\end{equation}
a scheme utilised by Amusia and Chernysheva in the \textit{hfgr} code \cite{atom_book}.


\begin{figure}[ht!]
\centering 
\includegraphics*[width=10cm]{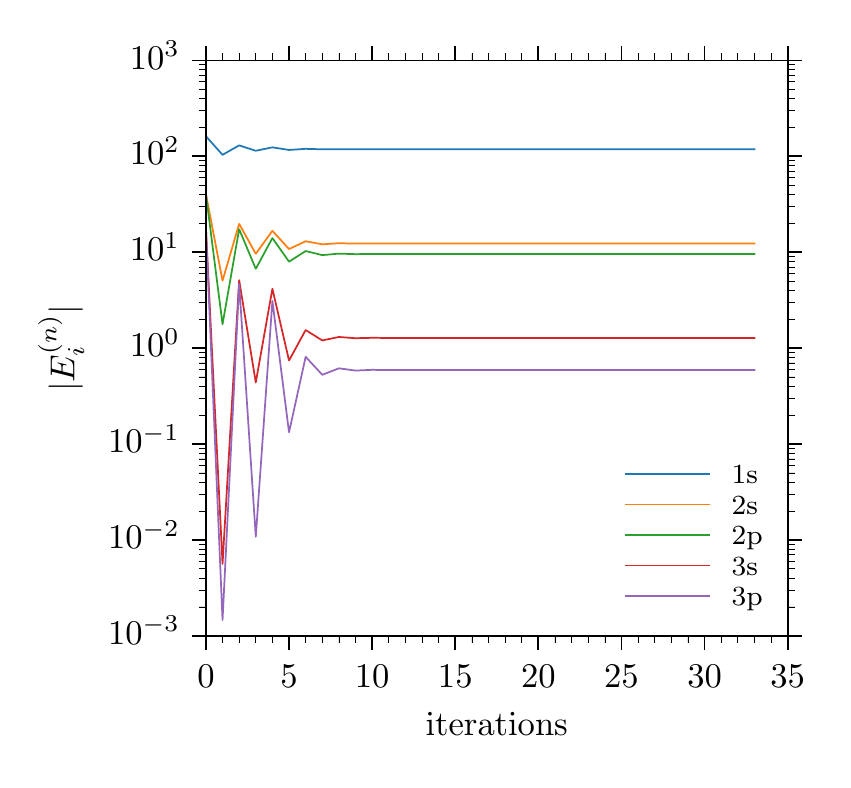}
\caption{Convergence of the occupied electron orbital energies for Ar against the number of iterations of the self-consistency process using $n=40$ splines of order $k=6$. No special techniques were necessary for the iterations to converge.}
\label{fig:Ar-noannealing}
\end{figure}

\begin{figure}[ht!]
\centering
\includegraphics*[width=10cm]{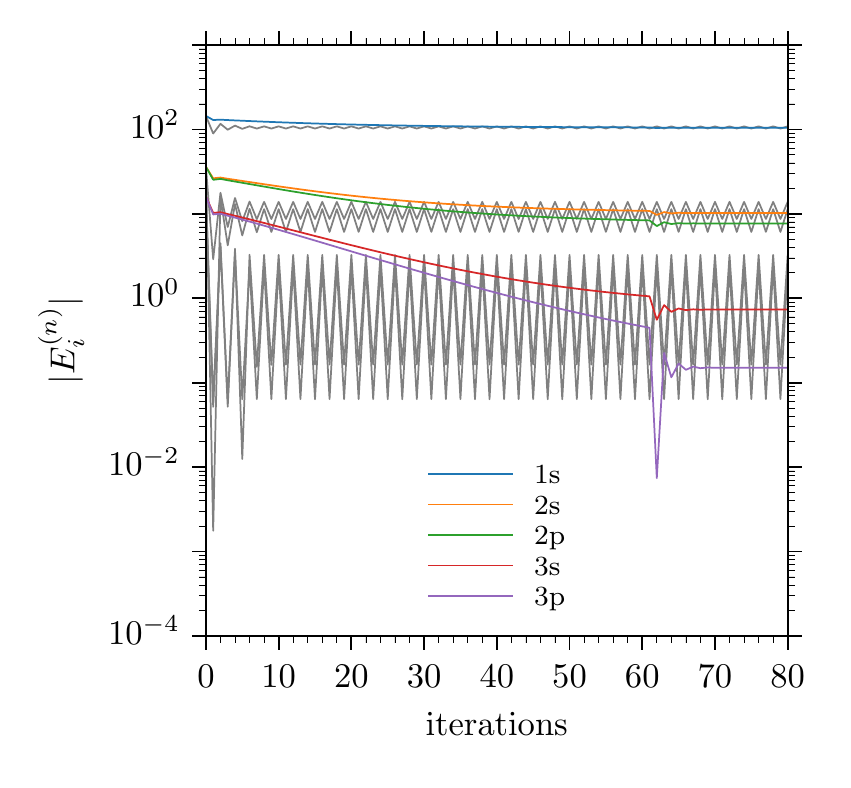}
\caption{Convergence of the occupied orbital energies for a negative ion, $\mathrm{Cl}^-$, with no `annealing' of the electron-electron interaction (grey), and the same system converged using annealing with parameters $\mathrm{e_s}=0.15$, $m_\mathrm{max}=50$ (colours).}
	\label{fig:Clm-test-annealing}
\end{figure}

Most closed shell neutral atoms converge without any special treatment. Figure~\ref{fig:Ar-noannealing} shows the progression of energies of occupied orbitals with iterations of the self-consistency process for Ar.
Negative ions are more unstable. 
Figure~\ref{fig:Clm-test-annealing} shows the analogous plot for the isoelectronic negative ion $\mathrm{Cl}^-$. The grey lines show that without additional measures, the system oscillates between two states and the iterations do not converge.
The coloured lines are also for Cl$^-$, but with the elementary charge gradually increased from 0.15 to 1.0. This is sufficient for achieving convergence for this system to the required $\varepsilon=10^{-12}$ threshold.

\begin{figure}[ht!]
\centering
\includegraphics*[width=10cm]{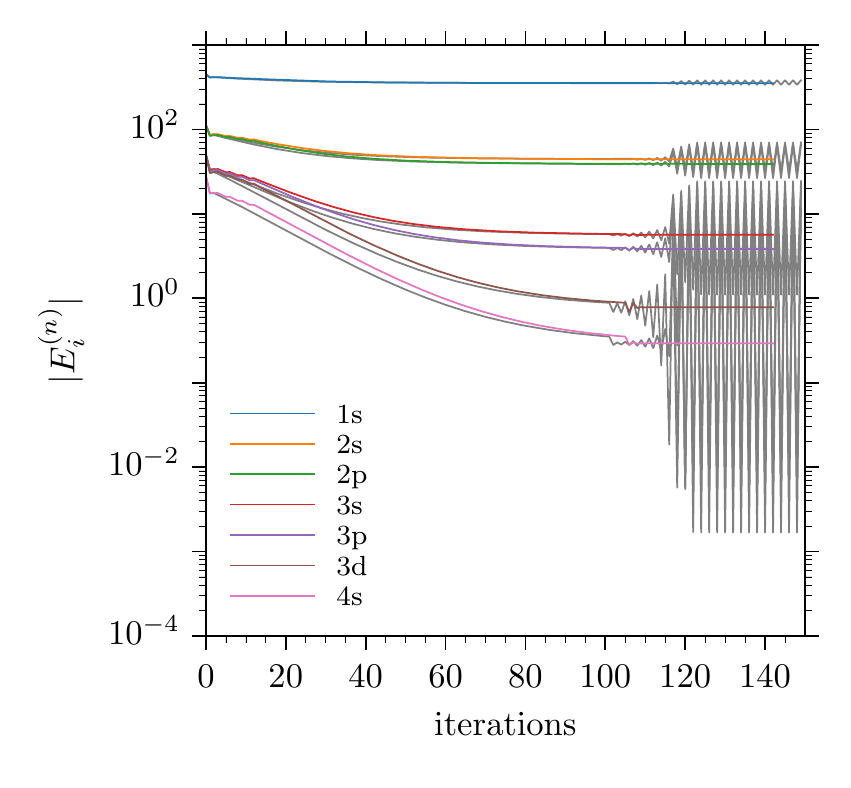}
\caption{Convergence of the electron orbital energies of Zn using `annealing' of the electron-electron interaction, Eq. (\ref{eq:anneal}) with $e_{\rm s}=0.9$, grey lines, and using `mixing' of the current and previous estimates of the solution, Eq.~(\ref{eq:wf_mix}) (coloured lines).}
\label{fig:Zn-test-mixing}
\end{figure}

One exception to the easy convergence displayed by for neutral atoms is Zn.
Figure~\ref{fig:Zn-test-mixing} shows that even using the electron-electron interaction `annealing', the solution becomes unstable beyond a certain point ($e_m\approx 1-10^{-5}$). This attempt is represented by the grey lines.
The coloured lines are for the same system but using a linear combination of the current and previous estimate of the self-consistent solution, Eq.~(\ref{eq:wf_mix}), which allows the iterations for Zn to converge.

The combination of these two techniques with sensible values of $e_\text{s}$ and $m_\text{max}$ allows rapid calculation of the ground- and excited- state energies and wavefunctions for a wide variety of systems, including neutral atoms and negative ions, as well as the electron gas confined by less singular, e.g., harmonic, potentials. 


\section{Results for neutral atoms} 
To demonstrate the robustness of the approach, we first consider neutral noble gas atoms. The computed ground-state energies of each orbital and the total energy are shown in Table \ref{tab:noble-gas-energies} with comparison to reference Hartree-Fock data by Saito \cite{saito_hartreefockroothaan_2009} and also experimental data \cite{radtsig_smirnov_1986,NIST_ASD}.
The calculations were performed using $n=40$ B-splines (in fact 38, as the first and the last B-spline are dropped, to satisfy the boundary conditions) of order $k=6$ to a self-consistent absolute error of $10^{-12}$. The tabulated values show that the calculated orbital energies are accurate to $\sim \! 10^{-6}$~a.u. with respect to the reference data.

\begin{table}[p]
\caption{\label{tab:noble-gas-energies}
Absolute values of the electron orbital energies and total ground-state energies for noble gas atoms calculated in the present work vs. reference Hartree-Fock calculations \cite{saito_hartreefockroothaan_2009} and experimental data \cite{radtsig_smirnov_1986,NIST_ASD}.}
\footnotesize
\lineup
\begin{tabular*}{\textwidth}{@{}l*{2}{@{\extracolsep{0pt plus 12pt}}d{4.9}}@{\extracolsep{0pt plus 12pt}}d{4.4}*{2}{@{\extracolsep{0pt plus 12pt}}d{5.9}}{@{\extracolsep{0pt plus 12pt}}d{4.3}}}
\br
 & \multicolumn{3}{c}{Energies (a.u.) for He, $Z=2$} & \multicolumn{3}{c}{Energies (a.u.) for Ne, $Z=10$} \\
& \crule{3} & \crule{3} \\
$nl$ & \mc{Present} & \mc{Ref.~\cite{saito_hartreefockroothaan_2009}} & \mc{Exp.} & \mc{Present} & \mc{Ref.~\cite{saito_hartreefockroothaan_2009} } & \mc{Exp.} \\
\mr
1s & 0.917955570 & 0.917956 & 0.904 & 32.772442840 & 32.772443 & 31.982 \\
2s & & & & 1.930390950 & 1.930391 & 1.781 \\
2p & & & & 0.850409731 & 0.850410 & 0.792^b \\
$E_\text{tot}$ & 2.861679994 & 2.861679996 & & 128.547098291 &  128.547098109 & \\
\mr
 & \multicolumn{3}{c}{Energies (a.u.) for Ar, $Z=18$} & \multicolumn{3}{c}{Energies (a.u.) for Kr, $Z=36$} \\
& \crule{3} & \crule{3} \\
1s & 118.610350525 & 118.610351 & 117.818 & 520.165468002 & 520.165468 & 526.508\\
2s & 12.322153702 &   12.322153 &  11.991 & 69.903082341  &  69.903082 &  70.669\\
2p & 9.571466070 &     9.571466 &   9.160 & 63.009785490  &  63.009785 &  62.315\\
3s & 1.277354057 &     1.277353 &   1.075 & 10.849466647  &  10.849466 &  10.768\\
3p & 0.591018393 &     0.591017 &   0.579 & 8.331501615   &   8.331501 &  7.977\\
3d & & &                                  & 3.825234561   &   3.825234 &  3.465\\
4s & & &                                  & 1.152935491   &   1.152935 &  1.011\\
4p & & &                                  & 0.524187023   &   0.524187 &  0.514\\
$E_\text{tot}$ & 526.817519726  &  526.817512803 & & 2752.054982444 & 2752.054977350 & \\
\mr
 & \multicolumn{3}{c}{Energies (a.u.) for Xe, $Z=54$} & \multicolumn{3}{c}{Energies (a.u.) for Rn, $Z=86$} \\
& \crule{3} & \crule{3} \\
1s & 1224.397777074 & 1224.397777 & 1270.093 & 3230.312837069 & 3230.312828 & 3616.134\\
2s &  189.340123037 &  189.340123 &  200.394 &  556.913115482 &  556.913115 &  663.436\\
2p &  177.782448960 &  177.782449 &  179.839 &  536.676971452 &  536.676971 &  570.411\\
3s &   40.175663257 &   40.175663 &   42.225 &  138.421866638 &  138.421866 &  164.637\\
3p &   35.221661899 &   35.221662 &   35.328 &  128.671558522 &  128.671558 &  137.504\\
3d &   26.118869411 &   26.118869 &   25.056 &  110.701350357 &  110.701350 &  108.043\\
4s &    7.856302172 &    7.856302 &    7.839 &   33.920746766 &   33.920746 &   40.241\\
4p &    6.008338645 &    6.008338 &    5.488 &   29.491183982 &   29.491183 &   31.114\\
4d &    2.777881328 &    2.777881 &    2.510 &   21.331318412 &   21.331318 &   20.102\\
4f &                &             &          &   10.107636171 &   10.107635 &    8.531\\
5s &    0.944414880 &    0.944414 &    0.860 &    6.905819457 &    6.905818 &    7.901\\
5p &    0.457290527 &    0.457290 &    0.446 &    5.225212748 &    5.225212 &    5.182\\
5d & & &                                     &    2.326320041 &    2.326319 &    1.838\\
6s & & &                                     &    0.873993818 &    0.873993 &    0.884^a\\
6p &  & &                                    &    0.428007511 &    0.428007 &    0.430^a\\
$E_\text{tot}$ & 7232.138377815 &  7232.138363870 & & 21866.772281217 &  21866.7722409 &\\
\br
\end{tabular*}
\noindent $^a$Energies from the NIST Atomic Spectra Database \cite{NIST_ASD}.

\noindent $^b$For orbitals with $l\geq 1$, the experimental energies are statistical averages of the fine-structure components, except for the outer $np$ orbital, where the value of the ionization potential is shown. 
\end{table}
\normalsize

Heavier atoms stretch the limits of the nonrelativistic HF approximation and improving the code to use the Dirac-Fock formalism should increase the performance in this regime. 
It can also be seen that, with the exception of the heaviest atom, radon, the results are within a few percent of the experimental ionization energy values.

\begin{figure}[ht!]
\begin{tikzpicture}
\draw (0,0) node[inner sep=0]{\includegraphics*[width=7.8cm]{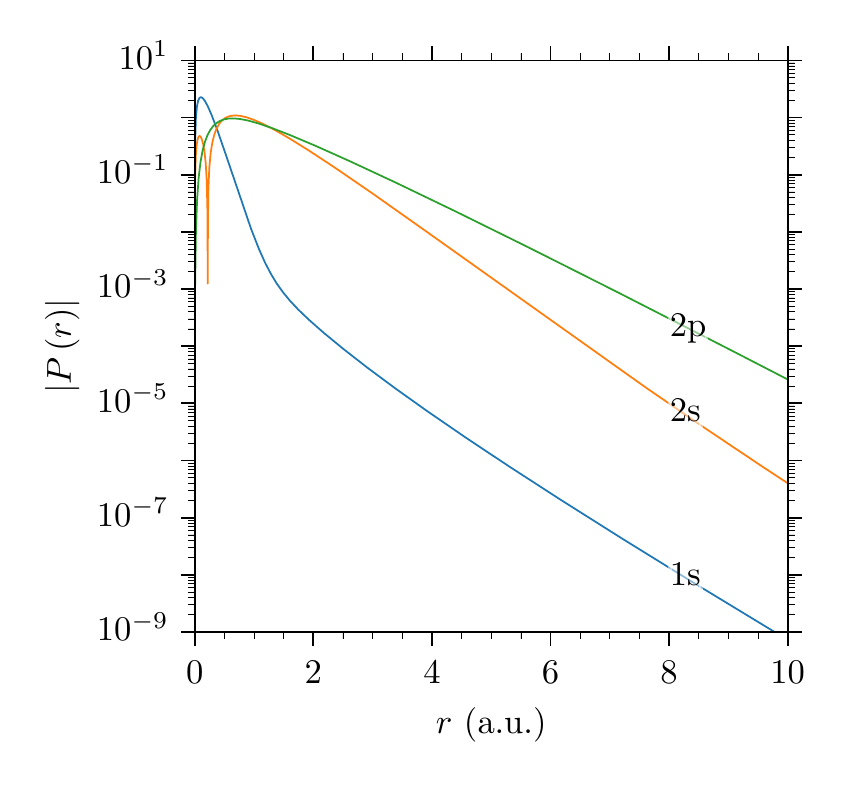}};
\draw (3.0,2.8) node{($a$)};
\end{tikzpicture}
\begin{tikzpicture}
\draw (0,0) node[inner sep=0]{\includegraphics*[width=7.8cm]{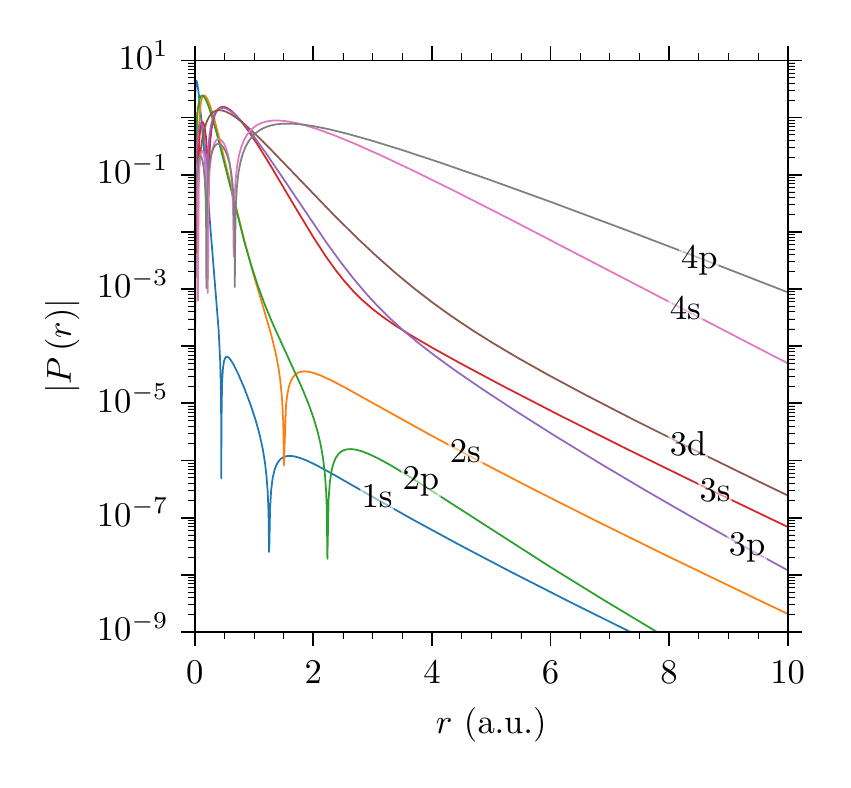}};
\draw (3.0,2.8) node{($b$)};
\end{tikzpicture}
\caption{The set of ground-state wavefunctions of (a) Ne, and (b) Kr, calculated using 100 B-splines of order 9, with $R=30\,\mathrm{a.u.}$ In ($a$) the wavefunctions have $n-l-1$ nodes, which appear as cusps when plotted on a log scale. However, in Kr ($b$) the inner orbitals exhibit extra nodes due to the exchange interaction \cite{dzuba_semiclassical_1982}.
}
\label{fig:example-wfns}
\end{figure}

Example sets of wavefunctions are shown in Figure \ref{fig:example-wfns} for Ne and Kr. Note that in both graphs the asymptotic behaviour of all the orbitals is similar to that of the outermost orbital, as expected from Handy \textit{et al.} \cite{handy_1969}, who found that the asymptotic behaviour of Hartree-Fock orbitals is proportional to the exponential of the orbital $i$ with the highest energy, i.e., the valence orbital,
\begin{equation}
P_j(r)\propto \exp\left(-\sqrt{2|E_i|}r\right).
\end{equation}
This is a consequence of the nonlocal exchange interaction between the electrons, as for a local potential the asymptotic behaviour would be $P_j(r)\propto \exp\left(-\sqrt{2|E_j|}r\right)$. Dzuba \textit{et al.} \cite{dzuba_semiclassical_1982} also noticed that the exchange interaction leads to additional nodes in some of the wavefunctions,
increasing them above the expected $n-l-1$ (i.e., the radial quantum number); see \cite{kozlov_exchange-assisted_2013} for futher background, additional references and physical implications. In Ne, shown in Fig.~\ref{fig:example-wfns} ($a$), the $1\mathrm{s}$ wavefunction has zero nodes, as expected. In Kr, Fig.~\ref{fig:example-wfns} ($b$), however, it has two nodes which show as downward cusps on the logarithmic scale. These extra nodes were considered ``undesirable'' in the past. However, arguments have been made recently that they may be responsible for observable effects such as the shapes of gamma-ray-spectrum peaks in positron-atom annihilation, the width of which is affected by annihilation on the inner shell electrons enhanced by exchange-assisted tunnelling \cite{kozlov_exchange-assisted_2013}.

\section{Results for harmonically confined electron gas: comparison between Hartree-Fock and the Thomas-Fermi model}\label{sec:harm-gas}

We now investigate the confinement of $N$ electrons in a harmonic potential
\begin{equation}
V(r) = \frac{1}{2}\omega^2 r^2,
\end{equation}
which is equivalent to a background field of constant positive charge density $\rho _b=\nabla ^2V/4\pi =3\omega ^2/4\pi $. This can be viewed as the simplest model for a metallic cluster, with $N$ free electrons in the background potential of $N$ singly-ionised (e.g., alkali) atoms distributed evenly in space (the ``jellium'' model).

We have calculated the orbital wavefunctions and energies for harmonic potentials with $\omega=1$ and $\omega=0.1$ a.u., for a series of closed-shell $N$-electron systems with $N \le 58$. In what follows we compare the results obtained using the Hartree-Fock approximation with those of the simpler Thomas-Fermi method, expected to be applicable for $N\gg 1$.

\subsection{Hartree-Fock calculations for harmonically confined electron gas}

\begin{figure}[ht!]
\begin{tikzpicture}
\draw (0,0) node[inner sep=0]{\includegraphics*[width=7.5cm]{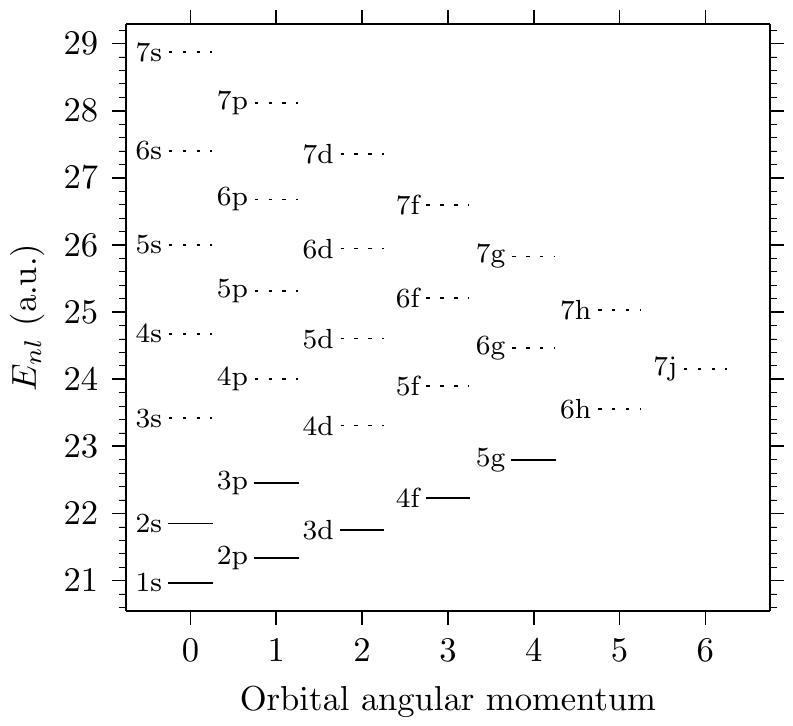}};
\draw (3.0,2.8) node{($a$)};
\end{tikzpicture}
\begin{tikzpicture}
\draw (0,0) node[inner sep=0]{\includegraphics*[width=7.5cm]{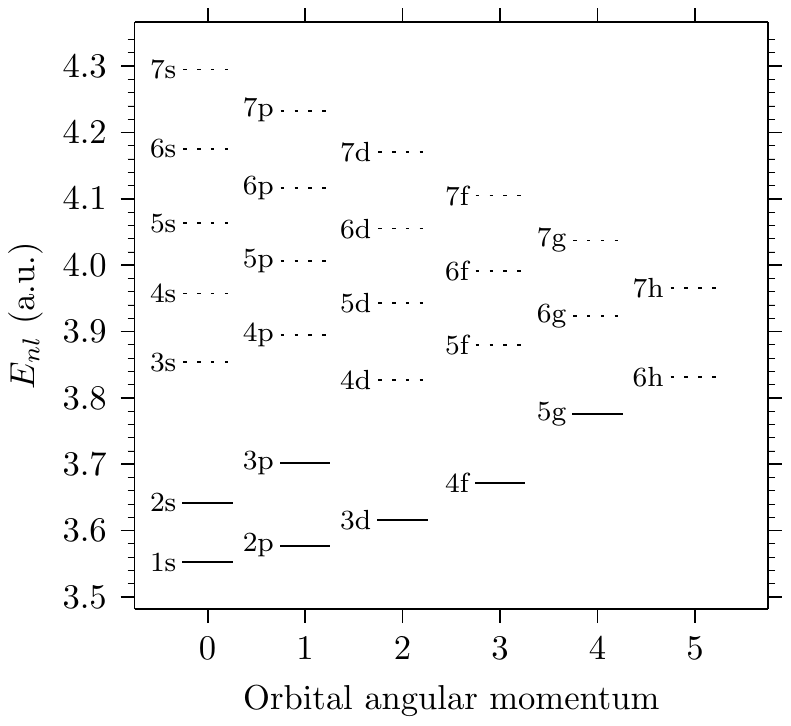}};
\draw (3.0,2.8) node{($b$)};
\end{tikzpicture}
\caption{The ground and exicted-state electron energy levels for a system of 58 electrons in a harmonic potential, ($a$) $\omega=1.0$~a.u., and ($b$) $\omega=0.1$~a.u.; solid lines indicate occupied orbitals, dotted lines --- excited states.}
\label{fig:harm-enes}
\end{figure}

The Hartree-Fock calculations were performed using the {\tt BSHF} code \cite{WaideCPC} with $n=40$ B-splines of order $k=6$. The ground-state orbital and total energies are shown in Tables \ref{tab:harm-1.0} and \ref{tab:harm-0.1}, in which successive columns show the energies for a system with an extra fully occupied orbital added, starting from the $1s^2$ configuration in the left-most column.
Note that for consistency we continue to use the hydrogenic orbital notation.
\begin{table}[ht]
\caption{\label{tab:harm-1.0} Energies of electron orbitals $nl$ and the total energy $E_\text{tot}$ for closed-shell systems with up to 40 electrons in a harmonic potential with $\omega=1$. Also shown is the number of iterations required to converge to an accuracy of $10^{-12}$.}
\footnotesize
\lineup
\begin{tabular*}{\textwidth}{@{}l@{\extracolsep{0pt plus 12pt}}d{1.6}@{\extracolsep{0pt plus 12pt}}d{2.6}*{4}{@{\extracolsep{0pt plus 12pt}}d{3.6}}}
\br
$N$ & \mc{2} & \mc{8} & \mc{18} & \mc{20} & \mc{34} & \mc{40} \\
$nl$ & \mc{$\text{1s}^2$} & \mc{$+ \text{2p}^6$} & \mc{$+ \text{3d}^{10}$} & \mc{$+ \text{2s}^2$} & \mc{$+ \text{4f}^{14}$} & \mc{$+ \text{3p}^6$} \\
\mr
\# iter.  & \mc{112} & \mc{116} & \mc{127}   & \mc{127}   & \mc{175}   &  \mc{177}\\
$E_\text{tot}$ & 3.771808  & 32.924181 & 121.395738 & 143.656801 & 340.876811 & 444.108657 \\
1s      & 2.259377  & 5.213001  &   9.168740 &   9.910487 &  14.372843 &  16.212245 \\
2p      &           & 6.015730  &   9.747907 &  10.464158 &  14.842209 &  16.599356 \\
3d      &           &           &  10.467192 &  11.143587 &  15.370154 &  17.084543 \\
2s      &           &           &            &  11.291091 &  15.520892 &  17.241711 \\
4f      &           &           &            &            &  16.007277 &  17.672330 \\
3p      &           &           &            &            &            &  17.886544 \\
\br
\end{tabular*}
\end{table}
\normalsize
\begin{table}[ht]
\caption{\label{tab:harm-0.1} Energies of electron orbitals $nl$ and the total energy $E_\text{tot}$ for closed-shell systems with up to 40 electrons in a harmonic potential with $\omega=0.1$. Also shown is the number of iterations required to converge to an accuracy of $10^{-12}$.}
\footnotesize
\lineup
\begin{tabular*}{\textwidth}{@{}l*{2}{@{\extracolsep{0pt plus 12pt}}d{1.6}}*{4}{@{\extracolsep{0pt plus 12pt}}d{2.6}}}
\br
$N$ & \mc{2} & \mc{8} & \mc{18} & \mc{20} & \mc{34} & \mc{40} \\
$nl$ & \mc{$\text{1s}^2$} & \mc{$+ \text{2p}^6$} & \mc{$+ \text{3d}^{10}$} & \mc{$+ \text{2s}^2$} & \mc{$+ \text{4f}^{14}$} & \mc{$+ \text{3p}^6$} \\
\mr
\# iter.  &   \mc{120}     &            \mc{126}   &              \mc{139}      &            \mc{139}     &             \mc{147}  &                \mc{154} \\
$E_\text{tot}$ & 0.529043  & 5.862360 & 23.198236 & 27.725662 & 67.695891 & 89.033498\\
1s      & 0.369424  & 1.097304 & 2.003531 & 2.164581 & 3.161704 & 3.552365 \\
2p      &           & 1.170797 & 2.044064 & 2.205478 & 3.193091 & 3.577094 \\
3d      &           &          & 2.113414 & 2.268952 & 3.233674 & 3.616198 \\
2s      &           &          &          & 2.292318 & 3.252050 & 3.641505 \\
4f      &           &          &          &          & 3.294960 & 3.672200 \\
3p      &           &          &          &          &          & 3.702001 \\
\br
\end{tabular*} 
\end{table}
\normalsize
Figure \ref{fig:harm-enes} shows the energies of the occupied and excited-state orbitals for a system with $N=58$ electrons for $\omega =1$ and $\omega =0.1$~a.u. The excited-state orbitals describe the states of an electron added to the system. The stronger confining harmonic potential with $\omega =1$ ~a.u. makes the energy-level spectrum closer to that of the harmonic oscillator.

\begin{figure}[ht!]
\begin{tikzpicture}
\draw (0,0) node[inner sep=0]{\includegraphics*[width=7.5cm]{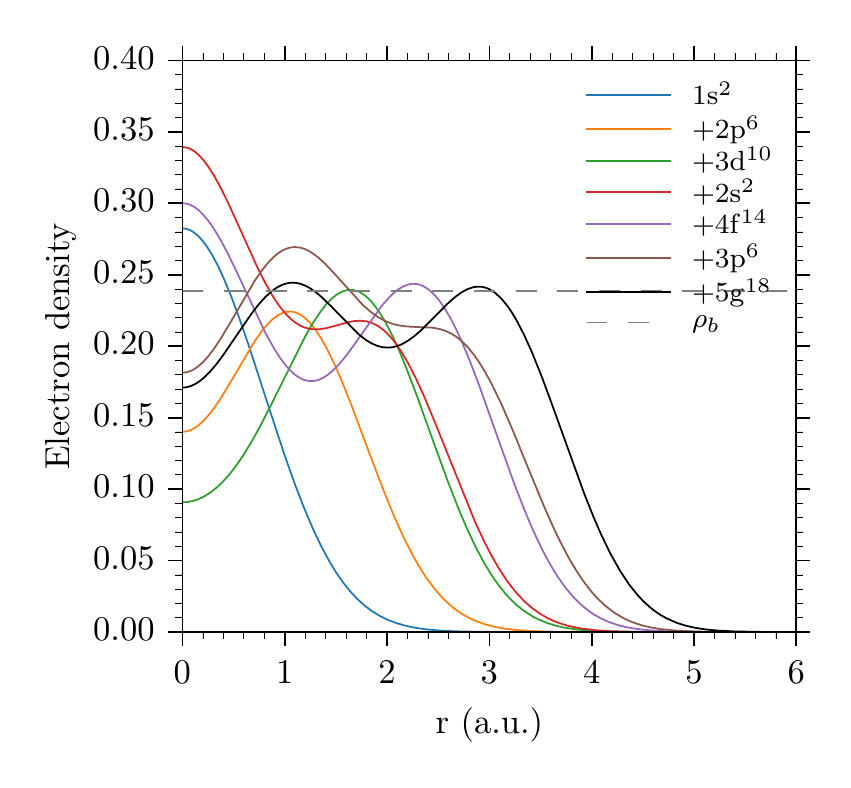}};
\draw (1.0,2.75) node{($a$)};
\end{tikzpicture}
\begin{tikzpicture}
\draw (0,0) node[inner sep=0]{\includegraphics*[width=7.5cm]{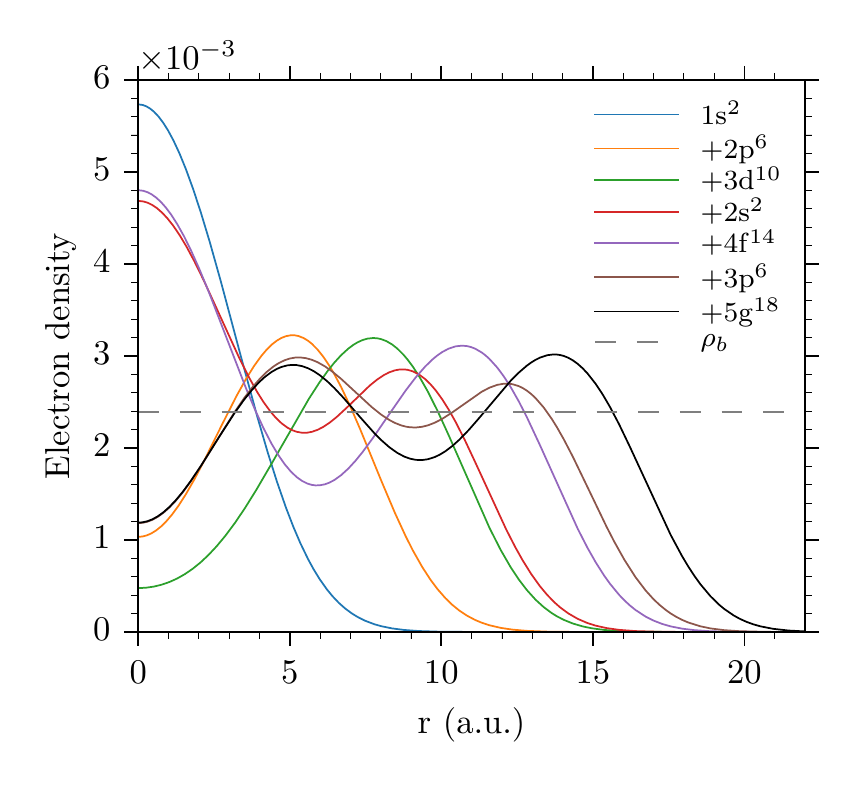}};
\draw (1.0,2.5) node{($b$)};
\end{tikzpicture}
\caption{The electron density for a harmonic well with an increasing number of electrons in closed shells for ($a$) $\omega=1.0$~a.u., and ($b$) $\omega=0.1$~a.u. Each successive line represents the density of the next closed-shell system, up to a total of 58 electrons.}
\label{fig:harm-density}
\end{figure}

Figure \ref{fig:harm-density} shows the charge density of closed-shell systems with $N=2$, 8, 18, 20, 34, 40 and 58 electrons. It can be seen that the radius of the system increases as more electrons are added and as the angular frequency ($\omega$) is reduced from 1~a.u. to 0.1~a.u. As electrons are added, the density becomes more uniform and closer to the positive background charge density $\rho _b$ shown by the grey dotted line. Indeed, as $N$ becomes large, the electron density acts to cancel out the positive background density giving the asymptotic relationship between $N$, $\omega$, and system radius $R$ as for a classical uniformly charged sphere, with $N\approx 4\pi R^3/3 \rho _b$, i.e., $R\approx N^{1/3}/\omega ^{2/3}$.

\subsection{Thomas-Fermi approximation}

To further test the validity of the results for the harmonically confined electron gas, we also analysed this system by adapting the well-known Thomas-Fermi method \cite{thomas_calculation_1927}, a semi-classical approximation for a system of electrons in which its state is represented by the electron density rather than a set of single-particle wavefunctions. It is often thought of as a precursor to the modern-day density functional theory, and although it fails to predict features of realistic systems, such as electronic shell structure, it is still often used since a qualitative view of the asymptotic behaviour of a system can be found comparatively easily \cite{brack_physics_1993,Flambaum21}.
This semi-classical approximation treats the electrons as occupying a density of states in phase space,
\begin{equation}
	\rmd {N_e}=2\cdot\frac{4\pi p^3}{3(2\pi)^3}\rmd {V},
\end{equation}
where a factor of $2$ has been included for spin degeneracy. Assuming that each of the electrons has a maximum single-particle energy given by,
\begin{equation}
	\varepsilon_0=\frac{1}{2}p_0^2+\frac{1}{2}\omega^2r^2-\phi_e,
\end{equation}
where $p_0$ is the maximum electron momentum,
the distribution of the electrons can be found from Poisson's equation,
\begin{equation}
	\nabla ^2(\phi_{\rm bg}+\phi_e)=4\pi(n_\mathrm{e}-\rho_{\rm bg}),\ n_\mathrm{e}=-\rho_\mathrm{e},
\end{equation}
where $\phi_\mathrm{\rm bg}$ and $\phi_\mathrm{e}$ are the harmonic background potential and the potential due to the charge distribution, $n_\mathrm{e}$. The harmonic potential is equivalent to a constant background charge density, $\rho_\mathrm{bg}=3\omega^2/4\pi$. Note that, in atomic units, the electronic charge distribution and equivalent potential differ by a factor of $-1$. 

Expanding the Laplacian in Poisson's equation, and assuming that the electron and background potentials are spherically symmetric, gives the non-linear differential equation,
\begin{equation}\label{eqn:tf-ode}
\frac{\rmd ^2 n_\mathrm{e}}{\rmd r^2}=
\frac{1}{3}n_\mathrm{e}^{-1}\left(\frac{\rmd n_\mathrm{e}}{\rmd r}\right)^2
- 2r^{-1}\frac{\rmd n_\mathrm{e}}{\rmd r}
+ \frac{4\pi}{B}n_\mathrm{e}^{4/3}
- \frac{3\omega^2}{B}n_\mathrm{e}^{1/3},
\end{equation}
where $B=(\pi^4/3)^{1/3}$. 

We take the electrons to be confined by the harmonic potential within a spherical box of radius $\mathrm{R_e}$ such that the electron charge density falls to zero everywhere outside this box.
Given that the harmonic potential tends to zero at the centre of the box, charge density of the electrons will approach a constant value, ${n_0}$, close to the centre, and at $r=0$ we assume that it is constant. 
These considerations constrain solutions of the ODE to satisfy the boundary conditions,
\begin{equation}\label{eqn:tf-bcs}
n_\mathrm{e}(0)=n_0,\quad 
\left.\frac{\rmd n_\mathrm{e}}{\rmd r}\right|_{r=0}=0,\quad 
n_\mathrm{e}(r)=0,\quad r\ge R_e.
\end{equation}
Integrating the charge density across the volume of the box must then give the total number of electrons in the system,
\begin{equation}\label{eqn:tf-norm}
4\pi\int_0^{R_e}r^2n_\mathrm{e}\rmd r=N_e.
\end{equation}
It can be readily seen from equation (\ref{eqn:tf-ode}) by substitution of the boundary conditions that for $n_e$ to decrease to zero then the initial charge density must satisfy $n_0<3\omega^2/4\pi=n_\text{bg}$. 

\begin{figure}
\centering
\includegraphics*[width=15cm]{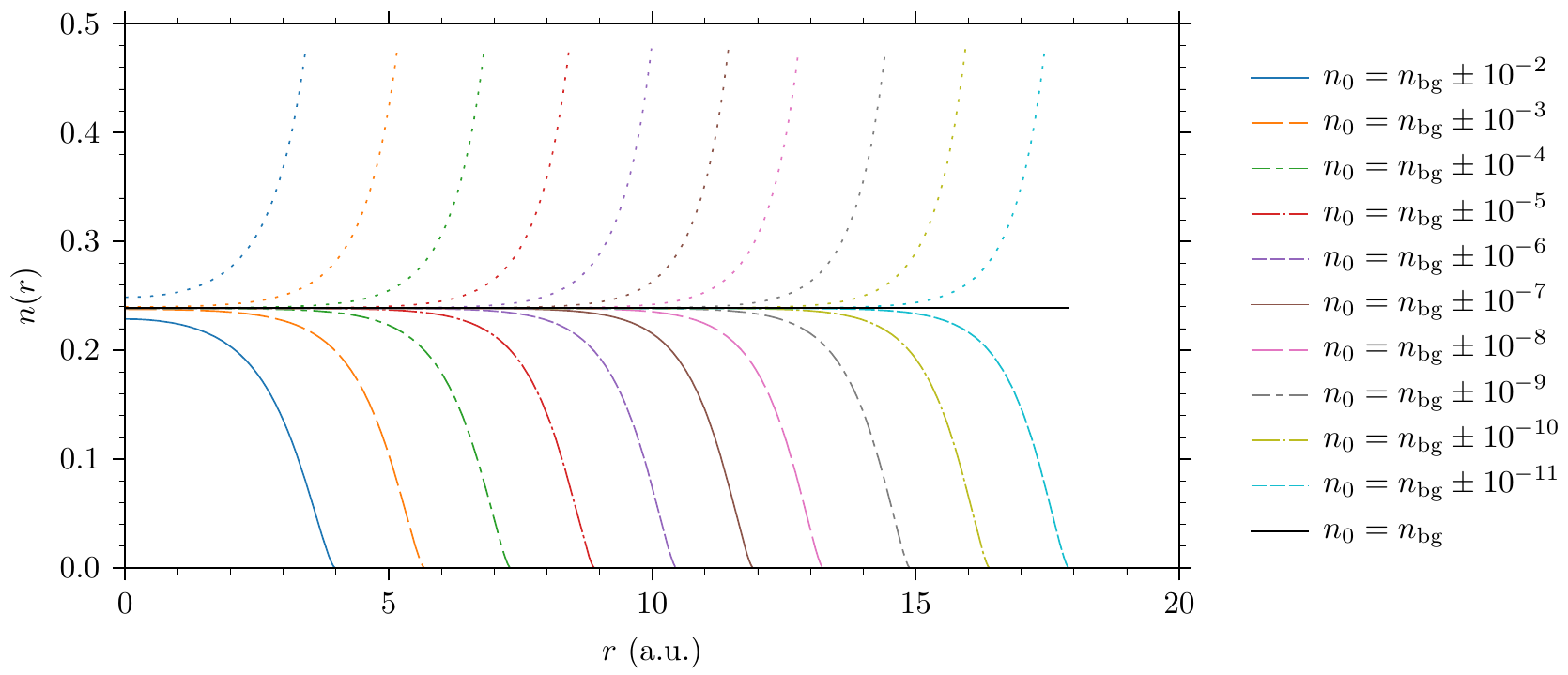}
\caption{Limits to the solution of the 1-D harmonic Thomas-Fermi equation ($\omega=1.0$). The black line is the background charge density $n_\text{bg}$, also the boundary between physical and unphysical solutions. Solutions with initial charge density $n_0 \ge n_\text{bg}$ diverge, whereas solution with $n_0 < n_\text{bg}$ reduce to $0$ at some point $r=R_e$.}
\label{fig:tf-sol-limits}
\end{figure}

Solutions to equation (\ref{eqn:tf-ode}), constrained by the conditions (\ref{eqn:tf-bcs}) and (\ref{eqn:tf-norm}) are shown in Figure (\ref{fig:tf-sol-limits}) for a range of values of the parameterised boundary condition, the initial charge density, $n_0$.
The boundary between physical and unphysical solutions, where $n_0=n_\text{bg}$ is shown as a solid black line.
Solutions with $n_0>n_\text{bg}$, shown as dotted lines, are monotonically increasing and therefore cannot represent systems with a finite number of electrons.
The physical solutions show that the size of the system $R_e$ increases approximately linearly with an order of magnitude decrease in the difference between $n_0$ and $n_\text{bg}$.
When solving the ODE numerically care must be taken to ensure that sufficient numerical precision is used. As $\omega$ decreases the required precision increases, e.g. for $\omega=0.1$, 40 significant figures are required to distinguish between systems with varying numbers of electrons.

To allow comparison with the Hartree-Fock results we need to be able to relate $n_0$ with the number of electrons in the system, $N$.
We solve the ODE for a range of $n_0$ values and calculate the corresponding $N$ by integration, as per equation (\ref{eqn:tf-norm}).
A non-linear function, described in section \ref{sec:tf-asymp-sol}, is then fitted to these points and used to compute $n_0$ for any desired $N$.

\subsubsection{Non-interacting solution.}
Equation (\ref{eqn:tf-ode}) is non-linear, and therefore likely intractable analytically.
However, a simple analytical solution may be found by ignoring the term arising from interaction between pairs of electrons, simplifying the equation.
This gives the electron number density as,
\begin{equation}\label{eqn:tf-ana-sol}
n(r)=\frac{1}{3\pi^2}\left(2\varepsilon_0-\omega^2r^2\right)^{3/2}.
\end{equation}
However, this is an over-simplification as the results deviate significantly from those of the full numerical solution. Comparing the magnitude of the energy of the interaction between pairs of electrons, $E_\mathrm{pot}^{(2)}$, with that of the harmonic potential energy, $V$, we can see that this approximation is only valid when $N\ll w$, i.e. as the electrons are more tightly bound their interaction with each other becomes less important. We do not consider this approximation further. 

\subsubsection{Asymptotic solution.}\label{sec:tf-asymp-sol}

Another way to arrive at an analytical expression for $n(r)$ is to consider its asymptotic form as $N_e$ becomes large.
Introducing the dimensionless function $\tilde{\eta}(r)$ defined by
\begin{equation}\label{eqn:density}
n(r)=n_\text{bg} \left(1-\frac{\tilde{\eta}(r)}{r}\right),
\end{equation}
and taking the ansatz
\begin{equation}
\tilde{\eta}(r)=A_1\exp(\beta r)+A_2\exp(\gamma r), \quad 
A_1,A_2,\beta,\gamma \in \mathbb{R},
\end{equation}
in Eqn.~(\ref{eqn:tf-ode}) results in the solution
\begin{equation}\label{eqn:tf-asym-sol}
\tilde{\eta}(r)=\frac{\eta_0}{\beta}\sinh(\beta r), 
\quad \eta_0=1-\frac{n_0}{n_\text{bg}}.
\end{equation}
From this, the asymptotic form for the radius is found to be  
\begin{equation}\label{eqn:tf-asym-fit}
R_e\simeq P+Q\ln( n_\text{bg}-n_0).
\end{equation}

Given that the electrons are trapped in a spherical box, it may be assumed that the mutual repulsion of the electrons leads to them being evenly distributed as their number becomes large. This makes the system similar to that of the uniformly charged insulating sphere. 
Therefore we take the relationship between the radius and the number of electrons to have the same form, $R_e \sim N_e^{1/3}$.

\subsubsection{Energies.}
For comparison with Hartree-Fock the total energy of the system, along with the kinetic, 1-body potential, and 2-body potential energies were calculated as follows.
\begin{equation}
E_\text{t} = E_\text{kin} + E_\text{pot}^{(1)} + E_\text{pot}^{(2)},
\end{equation}
\begin{equation}\label{eqn:tf-ene-kin}
E_\text{kin}=4\pi\chi\int_0^\infty n^{5/3} r^2 \rmd{r},
\quad \chi=\frac{3}{10}\left(3\pi^2\right)^{2/3},
\end{equation}
\begin{equation}\label{eqn:tf-ene-pot1}
E_\text{pot}^{(1)}=2\pi\omega^2\int_0^\infty n r^4 \rmd{r},
\end{equation}
\begin{equation}\label{eqn:tf-ene-pot2}
E_\text{pot}^{(2)}=\frac{1}{2}\int \!\!\!\!\int
\rmd ^3\bi{r} \rmd^3\bi{r}' \frac{n(r) n(r')}{|\bi{r}-\bi{r}'|}.
\end{equation}

\subsubsection{Fitting model.}

Electron number density curves $n(r)$ were computed for a range of values of $n_0$, distributed such that the values became more closely spaced as $n_0$ approached $n_\text{bg}$. The number of electrons represented by each solution was then calculated by integrating over $n(r)$. 

\begin{figure}
\centering
\includegraphics*[width=7.5cm]{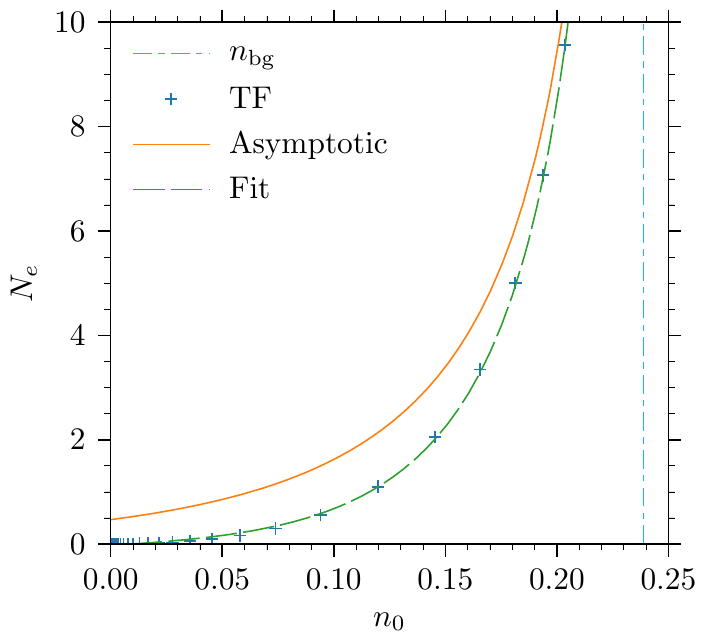}
\includegraphics*[width=7.5cm]{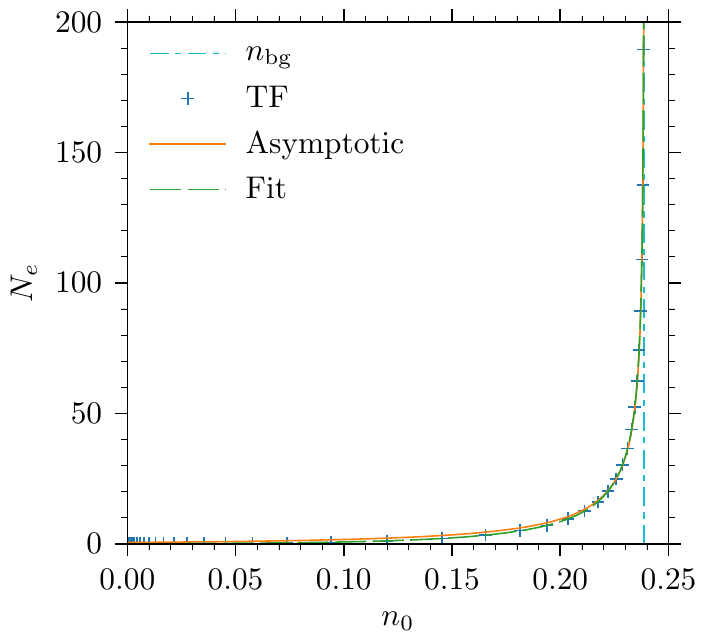}
\caption{
The number of electrons in the system, $N_e$ against the parametrized boundary condition, $n_0$ for a harmonic potential with $\omega=1$ in the Thomas-Fermi approximation with positive background charge density $n_{\rm bg}$ (vertical line). ``TF" is the Thomas-Fermi result (plus symbols), i.e., the discrete solutions of Eqns.~(\ref{eqn:tf-ode}--\ref{eqn:tf-norm}). ``Asymptotic" is the fit using expression Eqn.~(\ref{eqn:tf-asym-fit}) taking $R_e\sim N_e^{1/3}$ (orange line), and ``fit" that using Eqn.~(\ref{eqn:prag}) (green-dashed line).
}
\label{fig:tf-fit-w1_0}
\end{figure}

Figure (\ref{fig:tf-fit-w1_0}) shows $N_e$ plotted against the parameter $n_0$ for system with $\omega =1$. The dot-dashed line is the density corresponding to the positive background charge density $n_\text{bg}$.
The data has been fitted to the asymptotic behaviour function in Eqn.~(\ref{eqn:tf-asym-fit}). The same data has been shown at two different ranges for $N_e$, and it is clear from the figure that the function does fit the relationship for large values of $N_e$ (right panel) but not for smaller values (left panel). Indeed, in the limit as $n_0$ tends to zero, the number of electrons does not become zero. The relationship is almost quadratic for small $N_e$ so additional terms were added to the model to improve the fit for small $N_e$, giving
\begin{equation}\label{eqn:prag}
N_e \simeq \left[P+Q\ln(n_\text{bg}-n_0)\right]^3 + C_1 n_0^2 + C_2 n_0 + C_3,\quad C_i \in \mathbb{R},
\end{equation}
shown in Fig.~(\ref{fig:tf-fit-w1_0}) as the dashed line, the pragmatic fit. This now provides a good method of choosing the parameter $n_0$ corresponding to a system with a given number of electrons, $N_e$. This can be validated by integration once the solution, $n(r)$, is found.

\subsection{Comparison between Thomas-Fermi and Hartree-Fock}

\begin{figure}
\centering
\includegraphics[width=15cm]{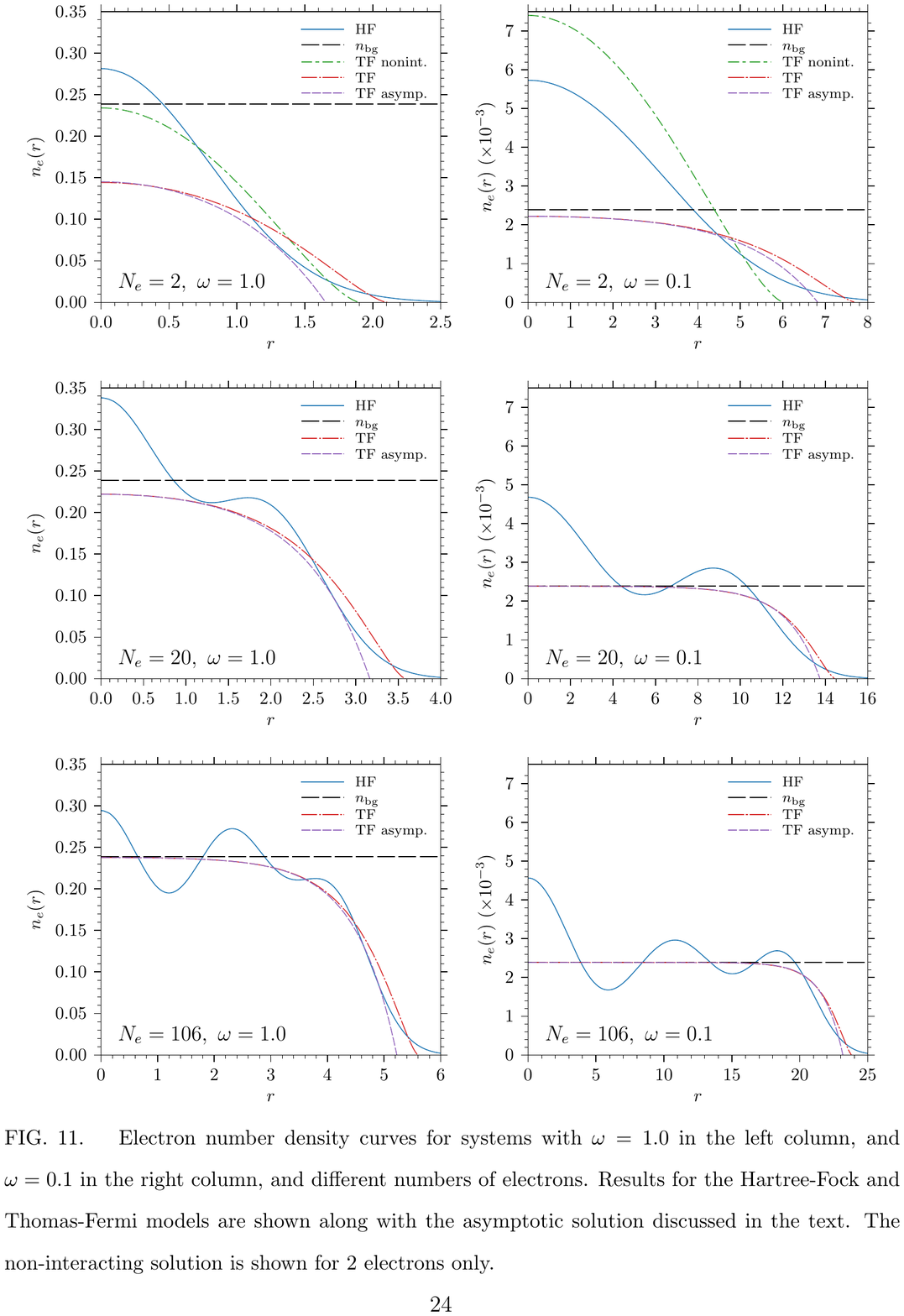}
\caption{Electron number density curves for systems with $\omega=1.0$ (left panels), and $\omega=0.1$ (right panels), and different numbers of electrons (indicated in bottom left of each panel). Equivalent background density $n_{\rm bg}=3\omega^2/4\pi$ (horizontal dashed line); Hartree-Fock result (blue solid line); Thomas-Fermi non-interacting model Eqn.~(\ref{eqn:tf-ana-sol}) (green dashed-dotted line in top left panel only), Thomas-Fermi result, i.e., the solution of Eqn.~(\ref{eqn:tf-ode}) (red dashed-dotted line), and Thomas-Fermi asymptotic result Eqn.~(\ref{eqn:density}) (purple dashed line).}
\label{fig:tf-orbs}
\end{figure}

In Fig.~(\ref{fig:tf-orbs}) the electron number densities, $n_\mathrm{e}(r)$, have been shown for systems with  $\omega$ (in the range 0.1--1) and $N_e$ (in the range 2--106). 
The radius of the system --- determined by the point where $n_\mathrm{e}$ falls to $10^{-3}\cdot n_\mathrm{e}(0)$ --- is dependent on $\omega$ and $N_e$,
\begin{equation}
R_e \propto \sqrt{\omega},\quad
R_e \propto N_e^{1/3},
\end{equation}
from comparison with the uniformly charged insulating sphere, and the behaviour of the system at the classical turning point. The plots give a visual comparison of HF calculated density with the results of the simpler Thomas-Fermi model. The analytic solution to TF, in which the electron-electron interaction has been ignored gives a fair approximation of the radius for $N_e=2$ but not for greater $N_e$, neither does it predict the form of the radial density curve. It has been ignored for larger systems.
The asymptotic solution in Eqn.~(\ref{eqn:tf-asym-sol}) becomes close to the form of the full numerical solution as $N_e$ becomes large.
Both predict the radius well and approximate the form of the HF solution at large $N_e$.
As $n_\mathrm{e}$ becomes uniform near the centre, the HF solution oscillates about the TF solution.
As the asymptotic and numerical solutions both become similar to HF for large $N_e$, we can be confident that the HF solutions are credible in this domain, i.e. that of finite numbers of fermions in a centrally-symmetric harmonic potential.
Further evidence supporting the similarity of the solutions is given in Table \ref{tab:tf-enes} which compares the kinetic, one-body, and two-body components of the total energy of different systems as calculated in both models. For all systems the energies of both models are broadly similar, increasingly so for large $N_e$ where they agree to within a few percent. 

\begin{table}[t]
\caption{\label{tab:tf-enes}The kinetic, one-body and two-body potential components of the total energy for systems with various $\omega$ and $N_e$, comparing the results of HF calculations with those of the Thomas-Fermi method.}
\footnotesize
\lineup
\begin{tabular*}{\textwidth}{@{}l@{\extracolsep{0pt plus 12pt}}d{3.0}*{4}{@{\extracolsep{0pt plus 12pt}}d{3.4}}*{2}{@{\extracolsep{0pt plus 12pt}}d{4.4}}}
\br
$\omega $ & N_e & \multicolumn{2}{c}{$E_\mathrm{kin}$}  & \multicolumn{2}{c}{$E_\mathrm{pot}^{(1)}$} & \multicolumn{2}{c}{$E_\mathrm{pot}^{(2)}$} \\
		 & & \mc{\text{HF}} & \mc{\text{TF}} & \mc{\text{HF}} & \mc{\text{TF}} & \mc{\text{HF}} & \mc{\text{TF}} \\
\mr
0.1 & 2 & 0.1071 & 0.0730 & 0.2121 & 0.2569 & 0.2098 & 0.3678 \\
0.1 & 20 & 1.0299 & 0.8939 & 9.5853 & 10.2809 & 17.1105 & 18.7742 \\
0.1 & 106 & 5.5175 & 5.0238 & 153.7706 & 157.474 & 296.5061 & 304.903 \\
\mr
1.0 & 2 & 1.3187 & 1.0819 & 1.7090 & 1.8124 & 0.7536 & 1.4611 \\
1.0 & 20 & 17.0992 & 15.7096 & 53.6006 & 55.2399 & 72.9687 & 79.0605 \\
1.0 & 106 & 101.8276 & 97.4406 & 767.4847 & 780.4820 & 1331.2867 & 1366.08 \\
\br
\end{tabular*}
\end{table}
\normalsize

\section{Polarizability of atoms and atomic clusters}

\subsection{Atoms}

A further test of the suitability of the wavefunctions was made by calculating the static dipole polarizability of the noble gas atoms 
\begin{equation}
\alpha = \frac{2}{3}2\sum_{nl,n'l'}
l_> \frac{|\langle n'l'|r|nl\rangle |^2}{E_{n'l'} - E_{nl}},
\end{equation}
where $l_> = \max\{l,l'\}$ and the sum $nl$ is over occupied orbital and $n'l'$ over excited orbitals. The results are given in Table \ref{tab:noble-stat-dip-pol} and are found to be in good agreement with results calculated using a similar Hartree-Fock method using the \textit{hfgr} code \cite{atom_book}, but in all cases the new B-spline values of {\tt BSHF} are closer to the tabulated reference values.

\begin{table}[ht]
\caption{\label{tab:noble-stat-dip-pol}Static dipole polarizability of neutral noble gas atoms calculated using a set of 40 B-splines of order 6 compared against the \textit{hfgr} code of Amusia \& Chernysheva and reference values \cite{atom_book,crc_handbook}.}
\begin{indented}
\item[]\begin{tabular}{@{}l*{2}{d{2.6}}d{2.7}}
\br
Atom & \mc{B-splines}  & \mc{hfgr}\cite{atom_book} & \mc{Ref. \cite{crc_handbook}} \\
\mr
He &  0.997236  &  0.997167 &  1.3837675 \\ 
Ne &  1.974636  &  1.973492 &  2.6717    \\ 
Ar & 10.140367  & 10.13132  & 11.0747    \\ 
Kr & 15.861810  & 15.84444  & 16.7656    \\ 
\br
\end{tabular}
\end{indented}
\end{table}

\subsection{Atomic Clusters}

Studying systems with multiple atoms would ideally be carried out using full quantum molecular calculations in which the nucleus of each atom and its associated electrons are all treated independently. Such calculations can be expensive. A (somewhat crude but inexpensive) approximation for clusters of atoms of a single element is to use the harmonic potential, as previously described in Sec.~\ref{sec:harm-gas}. 

The static dipole polarizabilities for closed-shell systems with $N$ electrons and a positive harmonic background potential for various values of $\omega$ are shown in Table \ref{tab:harm-pol-hf}. However, as the number of electrons in the system increases the effects of screening become more important, and this is underestimated in the HF approximation where electron correlations are not taken into account. The polarizabilites of the same systems were recalculated in the Random Phase Approximation (RPA), which includes a summation of electron-hole diagrams effectively describing the change in the electric field due to the interaction between electrons \cite{bohm_collective_1953}.
These results are shown in Table \ref{tab:harm-pol-rpae}.

\begin{figure}[t]
\centering
\includegraphics*[width=9cm]{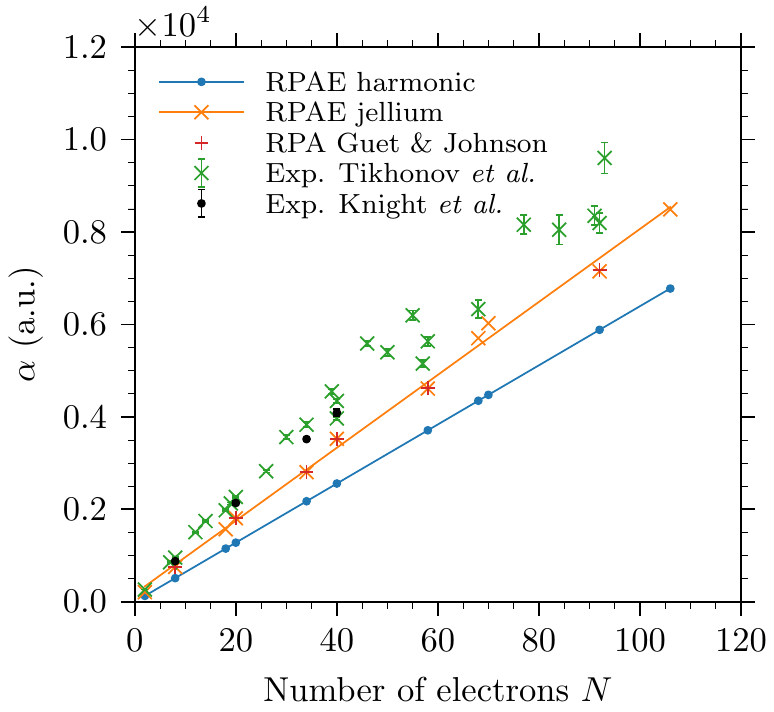}
\caption{Static dipole polarizability of a cluster of $N$ sodium atoms in the RPAE and jellium approximations benchmarked against similar calculations by Guet and Johnson \cite{guet_dipole_1992}. Data from two experiments, Tikhonov \textit{et al.} \cite{tikhonov_measurement_2001}, and Knight \textit{et al.} \cite{knight_electronic_1985}, and the results of the harmonic model are shown for comparison. Also see Table 8.}
\label{fig:jellium-rpa-comp}
\end{figure}

This harmonic model can be compared to experiment if we relate the parmeter $\omega$ to the radius of a physical system, in this case a cluster of $N$ sodium atoms; modelled, as previously, as a background of $N$ singly-ionised cations and $N$ independent electrons. Assuming the ions are evenly distributed in a sphere, we assign a radius of $R=N^{1/3}r_\mathrm{S}$, where $r_\mathrm{S}=4a_\mathrm{0}$ for Na.
This gives $\omega=0.125$; the static dipole polarizability for this system against the number of atoms, $N$, is shown in Fig.~\ref{fig:jellium-rpa-comp} (also see Table \ref{tab:jellium-rpa-comp}.) Experimental values from Knight \textit{et al.}~\cite{knight_electronic_1985}, and Tikhonov \textit{et al.}~\cite{tikhonov_measurement_2001} are also shown. The harmonic RPA calculations appear to follow a linear trend, similar to experiment but consistently underestimating $\alpha$ by $\sim $30\%. Note also that there is no evidence of shell structure shown in previous calculations \cite{de_heer_electronic_1987}.

\begin{table}[ht!]
\caption{\label{tab:harm-pol-hf}Static dipole polarizabilty in the Hartree-Fock approximation for electrons (closed electron orbitals up to and including that shown in the first column) in a harmonic well of frequency $\omega$.}
\footnotesize
\lineup
\begin{tabular*}{\textwidth}{@{}l*{3}{@{\extracolsep{0pt plus 12pt}}d{5.4}}*{2}{@{\extracolsep{0pt plus 12pt}}d{4.4}}*{2}{@{\extracolsep{0pt plus 12pt}}d{3.4}}}
\br
$N(nl)\backslash \omega $ & 0.1 & 0.125 & 0.2 & 0.4 & 0.6 & 0.8 & 1.0 \\
\mr
2 (1s$^{2}$) &  133.3161 &  87.1466 &  35.6765 &  9.5381 &  4.4028 &  2.5423 &  1.6599 \\
8 (+\,2p$^{6}$) &  998.9175 &  637.8120 &  247.2313 &  60.6943 &  26.5988 &  14.7960 &  9.3834 \\
18 (+\,3d$^{10}$) &  3244.2195 & 2056.7788 &  783.6019 &  186.3943 &  79.9463 &  43.7488 &  27.3808 \\
20 (+\,2s$^{2}$) &  3971.6733 &  2512.6102 &  952.2023 &  224.1820 &  95.3890 &  51.8094 &  32.1833 \\
34 (+\,4f$^{14}$) &  8237.8046 & 5205.9795 &  1967.3688 &  460.0682 &  194.6671 &  105.2793 &  65.1895 \\
40 (+\,3p$^{6}$) &  11526.4943 &  7257.9212 &  2714.9067 &  624.5585 &  261.7499 &  140.7227 &  86.8077 \\
58 (+\,5g$^{18}$) &  18721.1653 & 11772.8365 &  4408.5228 &  1015.7400 &  425.4560 &  228.4397 &  140.7094 \\
68 (+\,4d$^{10}$) &  25829.9106 & 16182.1026 &  6015.1382 &  1368.5965 &  568.4620 &  303.3674 &  186.0353 \\
70 (+\,3s$^{2}$) &  27636.3551 &  17281.3058 &  6406.1312 &  1452.8810 &  601.9823 &  320.4740 &  196.0164 \\
92 (+\,6h$^{22}$) &  38987.7679 & 24350.1336 &  9036.0681 &  2056.8043 &  853.4749 &  454.6450 &  278.1381 \\
106 (+\,5f$^{14}$) &  51155.2294 &  31752.8390 & 11704.7165 &  2647.6584 &  1093.6746 &  580.5109 &  354.0825 \\
\br
\end{tabular*}
\end{table}
\normalsize

\begin{table}[ht!]
\caption{\label{tab:harm-pol-rpae}Static dipole polarizabilty in the RPAE approximation for electrons (closed electron orbitals up to and including that shown in the first column) in a harmonic well of frequency $\omega$}
\footnotesize
\lineup
\begin{tabular*}{\textwidth}{@{}l*{1}{@{\extracolsep{0pt plus 12pt}}d{5.4}}*{2}{@{\extracolsep{0pt plus 12pt}}d{4.4}}*{4}{@{\extracolsep{0pt plus 12pt}}d{3.4}}}
\br
$N(nl)\backslash \omega $ & 0.1 & 0.125 & 0.2 & 0.4 & 0.6 & 0.8 & 1.0 \\
\mr
2 (1s$^{2}$) & 200.1095 & 128.0235 & 50.0818 & 12.5535 & 5.5955 & 3.1568 & 2.0263 \\
8 (+\,2p$^{6}$) &  799.9134 &  511.9628 & 199.9768 & 49.9847 & 22.2100 & 12.4900 & 7.9917 \\
18 (+\,3d$^{10}$) &  1799.6485 &1151.8284 & 449.9723 & 112.4934 & 49.9947 & 28.1202 & 17.9955 \\
20 (+\,2s$^{2}$) &  1999.6429 &  1279.8055 & 500.0243 & 125.0370 & 55.5871 & 31.2772 & 20.0238 \\
34 (+\,4f$^{14}$) &  3398.8268 &2175.3945 & 849.9711 & 212.5311 & 94.4721 & 53.1488 & 34.0207 \\
40 (+\,3p$^{6}$) &  3998.2571 & 2559.1648 & 999.8695 & 249.9788 & 111.0975 & 62.4889 & 39.9904 \\
58 (+\,5g$^{18}$) &  5796.3381 &3710.2519 & 1449.7408 & 362.4719 & 161.0975 & 90.6149 & 57.9917 \\
68 (+\,4d$^{10}$) &  6795.1135 &4349.5667 & 1699.6608 & 424.9729 & 188.8789 & 106.2433 & 67.9942 \\
70 (+\,3s$^{2}$) &  6994.9137 & 4477.1771 & 1749.6900 & 437.5068 & 194.4641 & 109.3949 & 70.0183 \\
92 (+\,6h$^{22}$) &  9191.0181 & 5884.4769 & 2299.3879 & 574.9845 & 255.5705 & 143.7665 & 92.0160 \\
106 (+\,5f$^{14}$) &  10588.4735 &6777.5321 & 2649.3053 & 662.4808 & 294.4547 & 165.6423 & 106.0156 \\
\br
\end{tabular*}
\end{table}
\normalsize

\subsection{Jellium model}

The harmonic model results can easily be improved by using the jellium model, rather than a constant positive harmonic background.
In the jellium model the harmonic background potential is limited to the defined radius of the system, $R$, and a Coulomb potential is used outside of this radius:
\begin{equation}
V(r) =
\cases{
-\frac{N}{2R}\left[ 3- \left( \frac{r}{R} \right)^2 \right] & $r \leq R$ \\
-\frac{N}{r} & $r > R$ 
}
\,.
\end{equation}
The harmonic model results in greater confinement of the electrons, thus reducing the value of $\alpha$. In the jellium model the electrons can more easily spill over beyond the system radius, $R$, increasing the polarizability and giving better agreement with experiment. Notably, the calculated values show evidence of shell structure and the corresponding computational results by Guet and Johnson are in excellent agreement with these new calculations \cite{guet_dipole_1992}.

\begin{table}[ht]
\caption{\label{tab:jellium-rpa-comp}Static dipole polarizabilites of a cluster of $N$ sodium atoms in the RPAE and jellium approximations, with similar calculations by Guet and Johnson \cite{guet_dipole_1992} and experimental results from Tikhonov \textit{et al.} \cite{tikhonov_measurement_2001}, and Knight \textit{et al.} \cite{knight_electronic_1985}. All polarizability values are in atomic units ($a_0^3$).}
\begin{indented}
\item[]\begin{tabular}{@{}ld{4.4}rrc}
\br
$N$ & \multicolumn{2}{c}{Calculations} & \multicolumn{2}{c}{Experiment} \\
& \mc{Present}  & \mc{Ref.~\cite{guet_dipole_1992}} & \mc{Knight \textit{et al.}} & \mc{Tikhonov \textit{et al.}} \\
\mr
2   &  208.7455 & --     & --               & $ 264.5 \pm   0.0$ \\
8   &  754.8100 &  755 & $ 879 \pm 17$ & $ 955.5 \pm   6.0$ \\
18  & 1569.7228 &  --    & --               & $1979.9 \pm  20.4$ \\
20  & 1807.8457 & 1808 & $2138 \pm 43$ & $2267.4 \pm  22.6$ \\
34  & 2805.5966 & 2806 & $3520 \pm 17$ & $3831.7 \pm  51.4$ \\
40  & 3525.5402 & 3529 & $4090 \pm 77$ & $3968.0 \pm  37.7$ \\
    	&           	&      	&               & $ 4345.9 \pm  37.7$\\
58  & 4613.1918 & 4619 &  --             & $5636.2 \pm  98.6$ \\
68  & 5702.8570 &  --    & --              & $6332.6 \pm 192.7$ \\
70  & 6026.9162 &  --    & --              & --\\
92  & 7149.6181 & 7178 & --              & $8195.2 \pm 208.6$ \\
106 & 8490.2581 & --     & --              & --\\
\br
\end{tabular}
\end{indented}
\end{table}

\newpage
\section{Summary and outlook}
An approach to the numerical solution of the Hartree-Fock equations that relies on gradually increasing the electron-electron interaction to its true value,  has been presented and found to give good numerical accuracy and fast and robust convergence. 
for neutral atoms, negative ions and electrons confined in harmonic potentials (for which comparison with the Thomas-Fermi model was made). 
The completeness of the manifold of excited-state wavefunctions was tested by calculating the static dipole polarizability of a range of neutral noble gas atoms in the static and Random-Phase Approximations, showing results consistent with previous methods. 
The HF basis states can be used in higher-order diagrammatic many-body calculations for systems with arbitrary central potentials, enabling calculations of electrons and positrons confined in electron gas \cite{makkonen_enhancement_2014}.

\vspace*{-1ex}
\section*{References}
\bibliographystyle{iopart-num}
\bibliography{schf2.bib}

\end{document}